\documentclass[%
 reprint, prl,
superscriptaddress,
 amsmath,amssymb,
 aps,
]{revtex4-2}
\usepackage{mathtools}
\usepackage{hyperref, soul}
\hypersetup{linktocpage,,colorlinks=true}
\usepackage{graphicx}
\usepackage{dcolumn}
\usepackage{bm}
\usepackage{physics}
\usepackage{tikz}

\begin{document}

\preprint{APS/123-QED}

\title{Impurity-induced non-unitary criticality}

\author{Heng-Hsi Li}
\email{steven0823255219@gmail.com} 
\affiliation{Department of Physics, National Tsing Hua University, Hsinchu 30013, Taiwan}
\author{Kuang-Hung Chou}%
\email{nagisa.7256.960247@gmail.com}
\affiliation{%
Department of Physics, National Tsing Hua University, Hsinchu 30013, Taiwan}%
\author{Xueda Wen}%
\email{xueda.wen@physics.gatech.edu}
\affiliation{%
School of Physics, Georgia Institute of Technology, Atlanta, GA 30332, USA}%
\author{Po-Yao Chang}%
 \email{pychang@phys.nthu.edu.tw}
\affiliation{%
Department of Physics, National Tsing Hua University, Hsinchu 30013, Taiwan}%
\affiliation{Yukawa Institute for Theoretical Physics, Kyoto University, Kyoto 606-8502, Japan}

\date{\today}

\begin{abstract}
Quantum impurities give rise to rich physical phenomena, with some exhibiting critical behavior described by conformal field theories (CFTs) in the low-energy limit. In parallel, party-time ($\mathcal{PT}$) symmetric non-Hermitian systems host exceptional points (EPs) at criticality, leading to exotic features governed by non-unitary CFTs. Here, we establish a connection between non-Hermitian impurities and CFTs by demonstrating that the critical properties of a (1+1)-dimensional free-fermion chain with central charge $c=1$
can be drastically altered by the presence of a local non-Hermitian impurity. Through a systematic analysis of entanglement/R\'enyi entropy, the finite-size scaling of the many-body spectrum, and fidelity susceptibility, we identify that this impurity-induced non-Hermitian criticality is characterized by a non-unitary CFT with central charge $c=-2$. Furthermore, we find that these non-unitary critical properties exhibit strong sensitivity to boundary conditions.
\end{abstract}

\maketitle


{\bf Introduction}---
Critical phenomena are among the most intriguing aspects of physics, characterized by drastic changes in physical quantities and the divergence of correlation lengths. At criticality, universality emerges, allowing distinct microscopic models to exhibit identical properties described by effective long-wavelength theories with conformal symmetries~\cite{DiFrancesco_book}.
In $(1+1)$-dimensional critical quantum systems, the Cardy-Calabrese formalism~\cite{Calabrese_2009,Calabrese_2004} predicts that the entanglement entropy $S_A$ of the subsystem size $l_A$ follows the universal scaling property,
$S_A = \frac{c}{3} \ln L_A + \rm{const.}$ with $c$ being the central charge reflecting the underlying degrees of freedom. 
This scaling behavior has been extensively validated in various quantum systems~\cite{Stephan_2014, Alcaraz2013,Vidal2003} and serves as a powerful tool for extracting the central charge of critical systems.

While most studies focus on Hermitian quantum systems, criticality in non-Hermitian quantum systems can also be described by non-unitary conformal field theories (CFTs)~\cite{Couvreur2017, chang2020entanglement, tu2022renyi}.
Unlike their unitary counterparts, which possess strictly positive central charges, non-unitary CFTs allow for negative central charges, leading to a negative logarithmic scaling of the entanglement entropy. Recent studies have demonstrated this property in various non-Hermitian models using a generalized entanglement/R\'enyi entropy adapted to non-Hermitian quantum systems via the biorthonormal basis~\cite{chang2020entanglement, tu2022renyi}.

A key feature of non-Hermitian systems is their extreme sensitivity to boundary conditions and defects. For instance, the non-Hermitian skin effect leads to the accumulation of a macroscopic number of boundary-localized states under open boundary conditions (OBC)~\cite{Lee2016, Leykam2017, Yin2018, Yao2018, Yokomizo2019, Lee2020}.
In such scenarios, the energy spectrum and eigenstates undergo drastic changes when transitioning from periodic boundary conditions (PBC) to OBC, a phenomenon absent in Hermitian systems. A similar effect occurs when a non-Hermitian defect is embedded in a Hermitian system, where even a single non-Hermitian impurity can induce scale-free localization of states~\cite{li2021impurity, Guo2023impurity}
Although this bears the resemblance to the non-Hermitian skin effect, it remains surprising that local non-Hermiticity can profoundly alters  properties of an originally Hermitian system.

In this work, we demonstrate that local non-Hermiticity can drive a system into non-unitary criticality described by non-unitary CFTs. Specifically, we consider a non-Hermitian impurity embedded in a one-dimensional critical free fermion chain. The non-Hermiticity respects parity-time (PT) symmetry, ensuring that the energy spectrum remains either entirely real or appears in complex conjugate pairs. The transition between the PT-preserving phase (real spectrum) and the PT-broken phase (complex spectrum) is characterized by an exceptional point (EP), where at least two energy eigenvalues and eigenstates coalesce.
We show that the presence of a local non-Hermitian impurity can induce an EP, leading to several hallmark properties of non-unitary CFTs. First, the entanglement entropy exhibits negative logarithmic scaling, with a central charge $c=-2$.
Second, the finite-size scaling of the many-body energy level spacing follows CFT predictions~\cite{Cardy_1984, CARDY1986} with $c=-2$.
Finally, using fidelity susceptibility, we explicitly identify the impurity-induced EP as the origin of this non-unitary criticality.


\begin{figure}  

\includegraphics[width=0.35\textwidth]{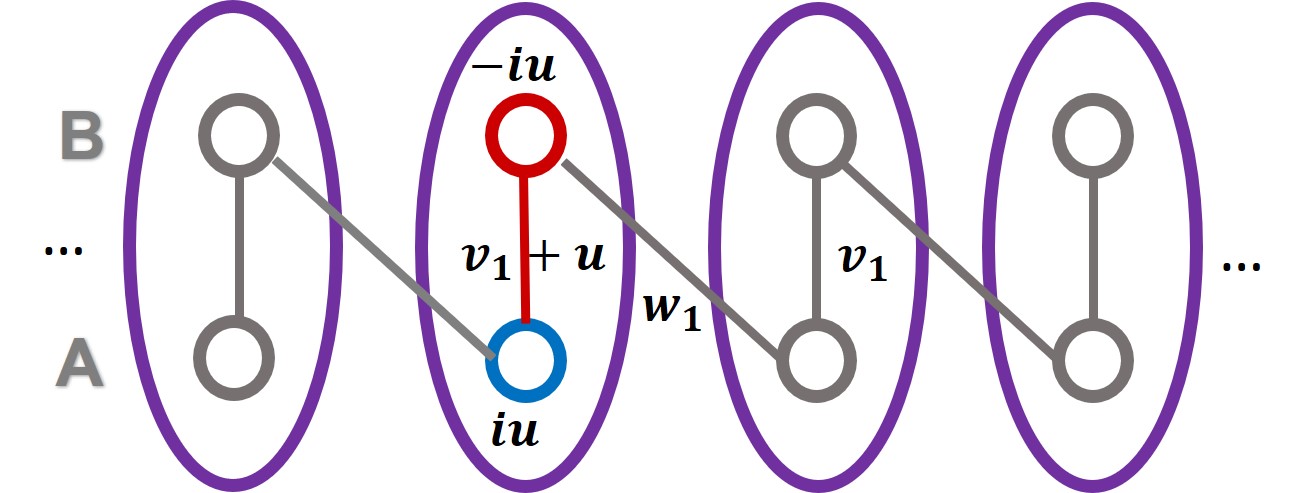}
  \caption{\label{fig:impuritymodel}
   The fermionic lattice tight-binding with the non-Hermitian impurity colored by red and blue, which correspond to on-site imaginary chemical potential $-i u$ and $+i u$ respectively.
  The hopping terms for the rest of the chain are alternating between $w_1$ and $v_1$. We consider  $v_1 =- w_1$ where the tight-binding chain is critical.}
\end{figure}

{\bf Critical chain with a non-Hermitian impurity}---We consider a Su-Schrieffer-Heeger (SSH) model with an non-Hermitian $\mathcal{PT}$-symmetric impurity described by the Hamiltonian $H_{\text{imp}}(\lambda) = H_{\text{SSH}} + V_{\text{imp}}(x, \lambda)$:
\begin{align} 
  \begin{split}
&H_{\text{SSH}} = \sum_{x = -L/2}^{L/2} (v_{1} c^{\dagger}_{B,x}c_{A,x} + w_{1} c^{\dagger}_{A,x+1}c_{B,x} + \text{h.c.} ), \\
& V_{\text{imp}}(x, \lambda) = (c^\dagger_{A,x}, c^\dagger_{B,x})
    \begin{pmatrix}
        iu\lambda^{2} & \lambda u \\
        \lambda u & -iu\lambda^{2}
    \end{pmatrix}\left(\begin{array}{c}c_{A,x} \\c_{B,x}\end{array}\right),
    \label{localperturbation}
  \end{split}
\end{align}
where $v_1(w_1)$ is the intra(inter)-hopping term, $(\lambda, u)$ are the parameters for non-Hermitian impurity which preserve the $\mathcal{PT}$ symmetry.
Here the $\mathcal{PT}$ symmetry maps $C_{A(B)x} \to C_{B(A)x}$ and $i \to - i$. Without loss of generality, we choose the coordinate $x \in [-L/2, L/2]$ and take the impurity at $x=0$.
We consider that the chain is critical by setting the hopping parameters $v_{1} = -w_{1}$, which leads to the gap closing point at $k = 0$ to avoid the parity effect.
For $\lambda=0$, the system is critical with the central charge $c=1$, where the central charge can be extracted from the entanglement entropy scaling $S_A = \frac{c}{3} \ln L_A + \cdots$ under periodic boundary condition (PBC),
and $S_A = \frac{c}{6} \ln L_A + \cdots$ under open boundary condition (OBC).  The entanglement entropy can be computed from the correlation matrix in free fermion models~\cite{chang2020entanglement}.
We first analyze the non-Hermitian impurity that influences the scaling property of the entanglement entropy by taking the impurity at $x=0$ with $\lambda =1$ as shown in Fig.~\ref{fig:impuritymodel}(b)

Although the impurity introduces non-Hermiticity to the critical chain, the system remains $\mathcal{PT}$-symmetric when $\lambda \in [0,1]$.
In the $\mathcal{PT}$ preserving region, the energy spectrum is real and the eigenvalues of the correlation matrix come in complex conjugate pairs, guaranteeing that the entanglement entropy remains real~\cite{chang2020entanglement}. 
Surprisingly, the local non-Hermiticity induced by the impurity can lead to the EP for the entire system which gives rise to drastic effects for several properties of the systems.
When the non-Hermitian impurity induced EP in the critical chain, the half-filling ground state includes the EP.
This ground state entanglement entropy exhibits a negative scaling with respect to the subsystem size~\footnote{In the ref \cite{chang2020entanglement}, they can achieve $c = -2$ scaling of EE since they do a shift momentum of the ground state, which the state nearby the EP can be filled. Based on this, they also examine that when EP is not filled, the EE scaling of the NHSSH model becomes $c = 1$}.

We solve the eigen-equations explicitly in this case~\cite{yao2018edge} and demonstrate that the EP can be induced by a single non-Hermitian impurity.
First, we assume that the EP state at $E=0$ induced by the non-Hermitian impurity at $x=0$  has the form $\phi_{x = 0} = (\phi_{A,0}, \phi_{B,0})^{T} \sim (1 , -i)^{T}$.
Second, we consider this EP state can be permeable to the Hermitian side by solving the eigen-equations 
\begin{align} \label{eigen}
    \begin{split}
    & (u + v_{1})\phi_{A,0} - iu \phi_{B,0} + w_{1} \phi_{A,1} = E\phi_{B,0} = 0, \\
    & v_{1}\phi_{B,1} + w_{1}\phi_{B,0} = E\phi_{A,1} = 0.
    \end{split}
\end{align}
We find that there is a solution for $\phi_{x \neq 0} =  (1, -i)^{T}$ which ensures the EP state permeates to the entire chain.
Interestingly, this solution exists as long as the Hermitian SSH chain is at criticality with PBC. The EP state becomes the eigenstate for the entire Hamiltonian.

\begin{figure*}[t]
  \includegraphics[width=0.95\textwidth]{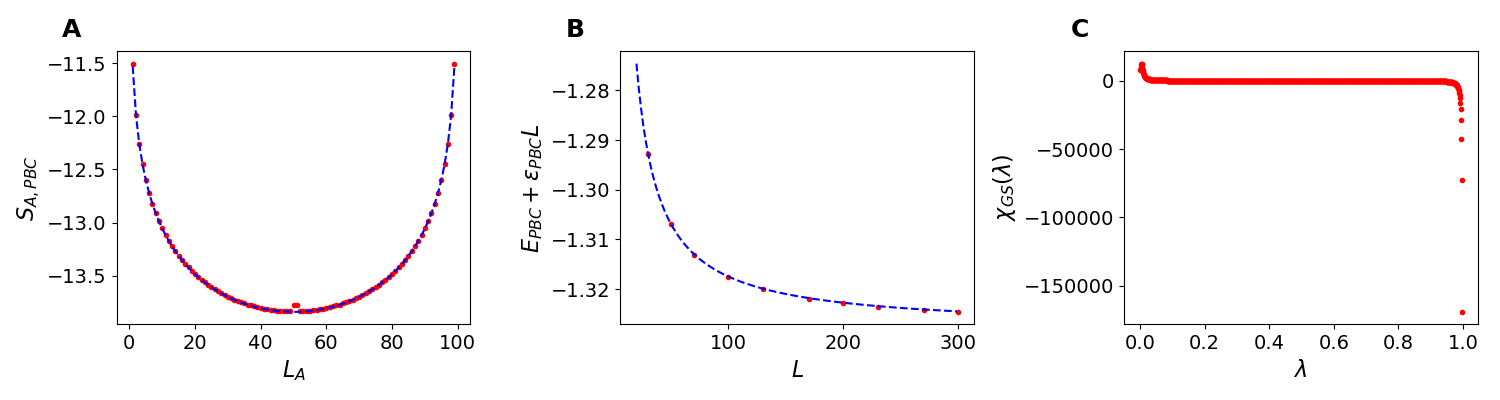}
  \caption{\label{fig:PBC} Entanglement entropy, Energy scaling, and many-body fidelity susceptibility under PBC. For the three figures, the parameters are $v_{1} = -w_{1} = 1$ ($v_{F} = 1$) and $u = 3$ (A) The entanglement entropy scales as $c = -2$ by the Cardy-Calabrese formula, the blue line is the fitting curve. (B) The parameters for the fitting curve (blue line) for the scaling \eqref{PBCMBE} are $\epsilon_{PBC} = 1.247239$, $A = -1.328$ and $C = 0.4117$. (C) For fidelity susceptibility, $\lambda$ runs from $0$ to $1$ by dividing it into $1000$ steps ,the fidelity susceptibility explodes when $\lambda \rightarrow 1$ and we pick $\epsilon \simeq 1/\text{steps} = 10^{-3}$ for simplicity.}
\end{figure*}

{\bf Probing the $c = -2$ non-unitary CFT  induced by non-Hermitian impurity}---
Once the ground state is permeated by the EP state from the non-Hermitian impurity, the many-body ground state is significantly changed.
We examine the properties of this EP-permeated many-body ground state from the entanglement entropy scaling, the finite-size energy scaling, and the many-body fidelity susceptibility. These properties indicate the system has similar behaviors as the $c = -2$ non-unitary CFT discussed in Ref.~\cite{chang2020entanglement}.

First, by performing bipartition on the lattice chain starting from $x = -L/2$, we examine the entanglement scaling under different subsystem sizes $L_{A}$ in Fig.\ref{fig:PBC} (A). The entanglement entropy scales as $c = -2$ according to the Cardy-Calabrese formula for PBC,
\begin{align} \label{PBCEE}
  S_{A} = \frac{c}{3}\log\bigg[\frac{L}{\pi \epsilon}\sin\bigg(\frac{\pi L_{A}}{L}\bigg)\bigg] + \text{const.}.
\end{align}
The origin of the negative scaling of the entanglement entropy is from the EP induced by the non-Hermitian impurity.
The EP state is exactly the same state as the EP state of the non-Hermitian SSH model at criticality, where the entanglement entropy scaling also gives the central charge $c=-2$~\cite{chang2020entanglement}.
We further examine the $n$-th R\'enyi entropy for the non-Hermitian impurity model, all the scaling properties of the  $n$-th R\'enyi entropy lead to $c=-2$~\cite{SM}.
This result indicates that the higher-order correlation function has the same behavior as the $c = -2$ non-unitary CFT.

In addition to the entanglement entropy scaling, we examine the finite size many-body energy scaling of the impurity model shown in Fig.\ref{fig:PBC}(B).
The scaling property satisfies the CFT description as  
\begin{align}\label{PBCMBE}
  E_{\text{PBC}}(L) = A + \epsilon_{PBC}L - \frac{\pi v_{F}}{6L}c + \frac{C}{L^{2}} + \dots,
\end{align}
where $A$ and $C$ are the constants related to the lattice details,
$\epsilon_{PBC}$ is the energy density, $v_{F}$ is the Fermi velocity which depends on the hopping strength, and $c$ is the central charge of the system. The scaling property of the many-body energy is shown in Fig.\ref{fig:PBC}(B).
The $L^{-1}$ term predicts the negative central charge $c = -2$ and serves as another evidence of $c = -2$ CFT induced by the local non-Hermitian impurity \eqref{localperturbation}.

Finally, we apply the many-body fidelity susceptibility $\chi_{GS}(\lambda)$~\cite{tu2022general}
to identify that the transition from $c=1$ to $c=-2$ is indeed induced by EP state with the non-Hermitian parameter $\lambda$.
The many-body fidelity susceptibility $\chi_{GS}(\lambda)$ is defined~\cite{tu2022general} as
\begin{align} 
  \begin{split}
F_{GS}(\lambda) &= \bra{GS_{L}(\lambda)}\ket{GS_{R}(\lambda + \epsilon)}\bra{GS_{L}(\lambda + \epsilon)}\ket{GS_{R}(\lambda )} \\
&= 1 - \chi_{GS(\lambda)}\epsilon^{2}+\mathcal{O}(\epsilon^3).
  \end{split}
\end{align}
where $\ket{GS_{R/L}(\lambda) }$ is the many-body ground state of the non-Hermitian Hamiltonian with parameter $\lambda$ in Eq.~(\ref{localperturbation}).
When the parameter $\lambda$ approaches the EP, the fidelity susceptibility goes to negative infinity $\chi_{GS(\lambda)} \rightarrow-\infty$~\cite{tu2022general}.
As shown in Fig.~\ref{fig:PBC}(C), the EP emerges at $\lambda \to 1$, where the system exhibits $c=-2$ non-unitary critical behaviors which can be identified from the finite-size many-body spectrum and the negative entanglement entropy scaling. One additional remark is that the above EP solution for PBC \eqref{eigen} is not limited to the case of single-impurity. One can add multiple impurities and the system still have EP solutions, resulting in the same conclusion on $c = -2$ for both entanglement entropy and many-body ground state energy scaling.
Our analysis demonstrates that a local non-Hermitian impurity can drive the critical system with $c=1$ to a non-unitary critical system with $c=-2$.

{\bf Other boundary conditions}---
While the above results are obtained under PBC, in non-Hermitian systems, boundary effects can be very drastic. For example, the non-Hermitian skin effect can lead to enormous states to be localized at the boundaries.
Here we investigate the scaling of the entanglement entropy and finite-size many-body spectrum under OBC and twist boundary condition (TBC).

For OBC, by considering the same perturbation as \eqref{localperturbation}, we calculate the entanglement entropy and observe that it has the scaling $c = 1$ entanglement entropy scaling. 
It is because the EP state disappears under OBC. To recover the EP state under OBC, we consider a generalized form of the $\mathcal{PT}$-symmetric impurity as 
\begin{eqnarray} \label{OBClperturbation}
  V_{\text{imp}}(0, \alpha, \beta) = 
    \begin{pmatrix}
        i\alpha\lambda^{2} & \lambda \beta \\
        \lambda \beta & -i\alpha\lambda^{2}
    \end{pmatrix},
\end{eqnarray}
where $\alpha > \beta$ to preserve $\mathcal{PT}$-symmetry for low energy states. We observe that when the parameters $(\alpha,\beta)$ are fine-tuned to host the EP state, the entanglement entropy still exhibits a logarithmic scaling property. However, the extracted central charge depends on $(\alpha,\beta)$.
For example, when $(\alpha,\beta)=(7/5,2/5)$, the entanglement entropy 
scales as 
\begin{align} \label{EEscaling}
  S_{A} \sim -\frac{2.798}{3}\log\bigg[\frac{L}{\pi \epsilon}\sin\bigg(\frac{\pi L_{A}}{L} \bigg)\bigg] + \text{const.} .
\end{align}
Here the configuration of the bipartition is that the subsystem $A$ contains one end of the chain.
The scaling property of the entanglement entropy from the Cardy-Calabrese formula under OBC, i.e. $S_{A} \sim \frac{c}{6}\log[\frac{L}{\pi \epsilon}\sin(\frac{\pi L_{A}}{L})] + \text{const.}$,
would lead to $c = -5.60$, which contradicts to the central charge $c = -2$ obtained from the PBC.
In addition to the entanglement entropy, we compute the $n$-th order R\'enyi entropy and observe the scaling property deviates from the logarithmic function for large $n$
[See Fig.~\ref{fig:renyiAppen}(F) in ~\cite{SM}].
For the finite size scaling of the many-body energy spacing under OBC, 
 the CFT predicts the form 
\begin{align}\label{OBCMBE}
  E_{\text{OBC}}(L) = A + \epsilon_{OBC}L - \frac{\pi v_{F}}{24L}c + \frac{C}{L^{2}} + \dots ,
\end{align}
where $A$ and $C$ are constants related to the lattice details, $\epsilon_{OBC}$ is the average energy for OBC system, $v_{F}$ is the Fermi velocity depending on the hopping strength and $c$ is the central charge of the system.
Again, the extrapolated central charge from the finite size scaling of the many-body energy spacing under OBC is  $c\sim-3.36$ [Fig.~\ref{fig:OBCfined}(B) in App.],
which does not agree with the result ($c=-2$) obtained from the PBC. 
Therefore, for fine-tuned parameters $(\alpha,\beta)$, the entanglement/R\'enyi entropy and the finite size scaling of the many-body energy spacing under OBC
does not have a consistent CFT description.

In addition to  PBC and OBC, we also analyzed the TBC by adding a local $U(1)$ gauge field $e^{i\phi}$ at the bond connecting the unit cells $x = -L/2$ and $x = L/2$. 
By fixing the coefficients at critical ($\phi=0$ with the $c=-2$ properties under PBC), we investigate the entanglement entropy scaling by changing the phase $\phi \in [0, 2\pi]$. We observe that as long as the phase $\phi$ deviates from $\phi = 0$ (PBC), the scaling of entanglement entropy goes to $c = 1$. This behavior can also be captured by the eigen-equations \eqref{eigen}, as there are no EP solutions when $\phi \neq 0$, which matches our observation of the relation between EP and the $c = -2$ entanglement entropy scaling. 
Our analysis demonstrates that the non-Hermitian impurity model is highly sensitive to boundary conditions. 

{\bf Discussion and final remark}--- 
In this paper, we investigate the effect of a  $\mathcal{PT}$-symmetric non-Hermitian impurity embedded in the free fermion critical chain.
Under PBC, we demonstrate that the EP can be induced by the non-Hermitian impurity which can be described by non-unitary CFT with $c=-2$ by using the logarithmic scaling of the entanglement/R\'enyi entropy and the finite size scaling of the many-body energy spacing. We identify the transition is explicitly governed by the EP from the many-body fidelity susceptibility.
This impurity induces non-unitary criticality is sensitive to the boundary conditions. A similar boundary sensitivity has also been observed in Yang-Lee modeml where the spectrum changes quite
abruptly with boundary condition variations~\cite{Gehlen_1991}.  
Additionally, the entanglement entropy scaling in complex CFTs also suggests that the boundary effects play a subtle role in determining the central charge~\cite{Shimizu2025},  indicating that the boundary phenomena in non-unitary/complex CFTs are rich and merit further investigation.
Finally, our model might be interpreted as a non-unitary analog~\cite{Castro-Alvaredo_2017, Xu2022} of the defect C-theorem~\cite{Nishioka2021,Sato2021,Kobayashi2019,Carreno2023,Bolla2023}, where the central charge flows from $c=1$ to $c=-2$ due to the local non-Hermitian perturbation.


\section{\label{sec:acknowledgments} Acknowledgments}
\begin{acknowledgments}
We would like to thank Chang-Tse Hsieh for his insightful discussion. P.-Y. C. acknowledges support from the National Science and Technology
Council of Taiwan under Grants No. NSTC 113-2112-M-
007-019 and the support from Yukawa Institute for Theoretical Physics, Kyoto University. 
X. W. is supported by a startup at Georgia Institute of Technology.
H.-H. L., K.-H. C. and P.-Y. C. thank the National
Center for Theoretical Sciences, Physics Division for its support.
\end{acknowledgments}

\bibliography{nHRef}

\clearpage

\pagebreak

\newpage

\setcounter{equation}{0}
\setcounter{figure}{0}
\setcounter{table}{0}
\setcounter{page}{1}
\makeatletter
\renewcommand{\theequation}{S\arabic{equation}}
\renewcommand{\thefigure}{S\arabic{figure}}
\renewcommand{\bibnumfmt}[1]{[S#1]}
\renewcommand{\citenumfont}[1]{S#1}

\begin{widetext}
\begin{center}
\noindent{\Large Supplementary Material for ``Impurity-induced non-unitary criticality". }
\end{center}

In this document we present the finite-size scaling of the many-body spectrum, many-body fidelity susceptibility, and entanglement entropy for the single impurity model under the open boundary condition (OBC).
We also analyze the effects of the fine-tuned parameters under periodic boundary condition (PBC).

\section{\label{sec:appenA} Additional analysis under OBC}
Before diving into the OBC part, we provide a short check for the many-body ground state energy scaling PBC SSH model in Fig.\ref{fig:nHPBC}. Notice that our result is different from the conclusion from the recent paper \cite{zhou2024universal}, in which they claim that the PBC will possess $c = 1$ under energy scaling. 
Alternatively, we get the $c = -2$ energy scaling that matches the property of the EP wavefunctions as argued in \cite{chang2020entanglement}.

\begin{figure}[ht]
  \includegraphics[width=0.46\textwidth]{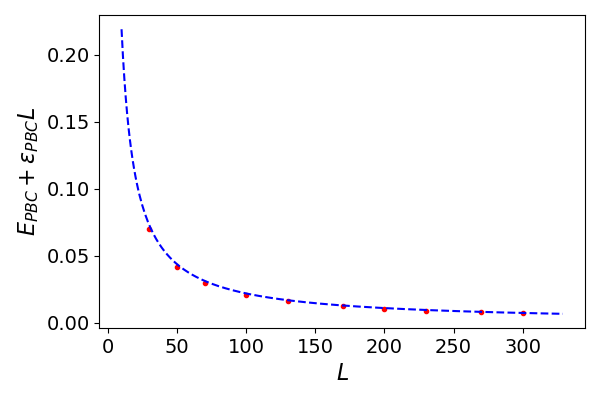}
  \caption{\label{fig:nHPBC} Many-body ground state energy scaling for non-Hermitian SSH model with parameters $v_{1} = -w_{1} = 1$ (in nHSSH is $v = 4$, $w = -1$ and the corresponding Fermi velocity $v_{F} = 2$) and $u = 3$, and for the fitting curve (blue line)\eqref{PBCMBE} are $\epsilon_{PBC} = 2.546479$, $A = - 3.13\times 10^{-6}$ and $C = 2.546479$. The corresponding central charge is $c = -2$ which is different from the conclusion of Ref.\cite{zhou2024universal}.}
\end{figure}

\begin{figure*}[ht]
  \includegraphics[width=0.95\textwidth]{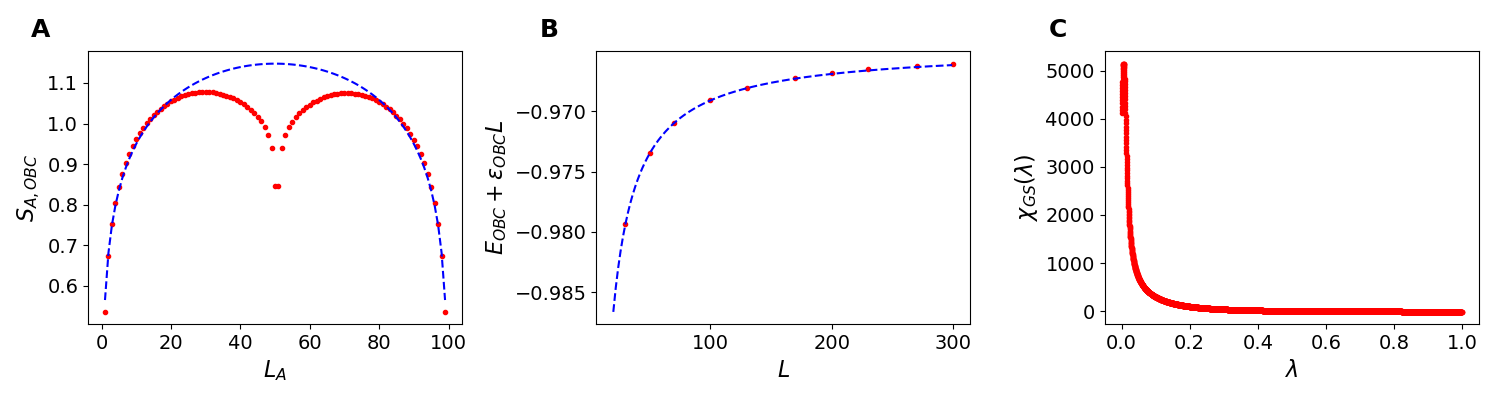}
  \caption{\label{fig:OBC} Entanglement entropy, Energy scaling, and many-body fidelity susceptibility under OBC with perturbation \eqref{localperturbation}. For the three figures, the parameters are $v_{1} = -w_{1} = 1$ ($v_{F} = 1$) and $u = 3$ (A) The entanglement entropy scales as $c = 1.01$ by the Cardy-Calabrese formula. For the fitting curve (blue line), we ignored the short-range entanglement effect caused by the impurity. (B) The parameters for the fitting curve (blue curve) for the scaling \eqref{OBCMBE} are $\epsilon_{OBC} = 1.27324$, $A = -0.9647$, $C = 0.2827$, and the extracted central charge is $c = 1$  (C) Many-body fidelity susceptibility where $\text{steps} = 4000$. One can decrease the initial fidelity susceptibility with order $O(10^{3})$ by choosing different paths of $\mathcal{PT}$-symmetric perturbation.}
\end{figure*}

\begin{figure*}[t]
  \includegraphics[width=0.95\textwidth]{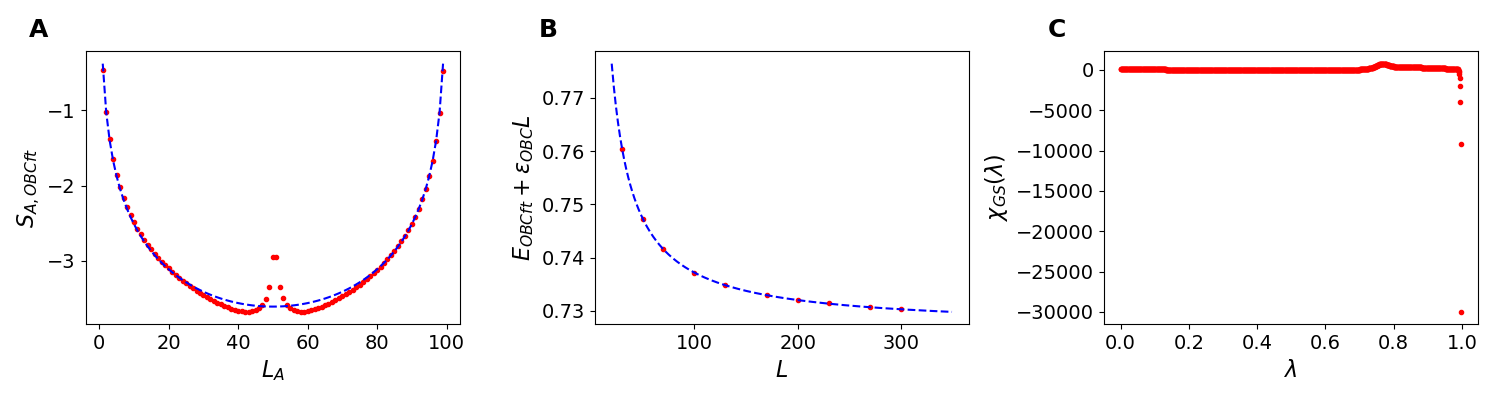}
  \caption{\label{fig:OBCfined} Entanglement entropy, Energy scaling, and many-body fidelity susceptibility under OBC with fine-tuned impurity \eqref{OBClperturbation}. For the three figures, the parameters are $v_{1} = -w_{1} = 1$ ($v_{F} = 1$) and $\alpha = 7/5$ and $\beta = 2/5$ for impurity. (A) The entanglement entropy scales as $c = -2.798$ by the Cardy-Calabrese formula (the blue fitting curve), and there are short-range entanglement effects due to the impurity. (B) The parameters for the fitting curve (blue line) for the scaling \eqref{OBCMBE} are $\epsilon_{OBC} = 1.27324$, $A = 0.7268$, and $C = -1.111$. However, the parameters will give a different central charge $c \simeq -3.36$, compared to the entanglement entropy. (C) Many-body fidelity susceptibility where $\text{steps} = 1000$.}
\end{figure*}

First, we start with local perturbation \eqref{localperturbation} under OBC. In Fig.\ref{fig:OBC}(A), the entanglement entropy tells a $c = 1$ scaling. The short remark here is that the short-range effect can be removed by changing the parameters $u$. As for the many-body energy scaling in Fig.\ref{fig:OBC}(B), the energy scaling gives $c = 1$ which agrees with the entanglement entropy. Also, the many-body fidelity susceptibility in Fig.\ref{fig:OBC}(C) does not diverge as $\lambda \rightarrow 1$. Mention here that the initial value of the many-body fidelity susceptibility is caused by the finite steps of the numerics, as the initial value may decrease whenever the step increases. 

Then, for the fine-tuned impurity model \eqref{OBClperturbation}, we analyze the entanglement entropy in Fig.\ref{fig:OBCfined}(A) which gives a $c \sim -2.789$ scaling caused by the transition to EP as mentioned in the main text. However, the many-body energy scaling yields a different central charge $c\sim-3.36$, providing evidence that the negative scaling arises from the EP. Also, similar to the previous cases, we also look at the many-body fidelity susceptibility Fig.\ref{fig:OBCfined}(C), giving a similar diverging behavior as Fig.\ref{fig:PBC}. Supporting our observation that the EP is closely related to the negative scaling of entanglement entropy.

\section{\label{sec:appenB} R\'enyi entropies for PBC impurity models and OBC fine-tuned parameters}
The n-th order R\'enyi entropy $S^{(n)}_{A}$ is given by \cite{tu2022renyi}, in which they build the modified definition on the correlation matrix
\begin{align} \label{RenyiEE}
  \begin{split}
  S^{(n)}_{A, PBC} 
  &= \frac{c(n + 1)}{6n} \log\bigg[\frac{L}{\pi \epsilon}\sin (\frac{\pi L_{A}}{L})\bigg] + \text{const.}, \\
  S^{(n)}_{A, OBC} 
  &= \frac{c(n + 1)}{12n} \log\bigg[\frac{L}{\pi \epsilon}\sin (\frac{\pi L_{A}}{L})\bigg]+ \text{const.}.
  \end{split}  
 \end{align}
which reduces to the entanglement entropy as $n \rightarrow 1$. Such R\'enyi entropy is argued to be related to the higher-order correlation function in the field theory and can serve as the check whether the theory obeys the prediction from the CFT.

Here, we make a short check if the PBC together with OBC possess similar $c = -2$ CFT scaling under the R\'enyi entropies \eqref{RenyiEE}. In Fig.\ref{fig:renyiAppen} (A,C,E), we show that the impurity model under PBC has the scaling following the formula \eqref{RenyiEE}. This result implies us that the PBC impurity model may have the identical CFT structure as mentioned in \cite{chang2020entanglement}, which implies that the central charge changes from $c = 1$ to $c = -2$ induced by local non-Hermitian impurity as mentioned in the main text. However, for the fine-tuned OBC impurity model in Fig.\ref{fig:renyiAppen}(B,D,F) with $\eqref{OBClperturbation}$, the higher-order R\'enyi entropies such as the second R\'enyi entropy $S_{A}^{(2)}$ fails to follow the logarithmic scaling~\eqref{RenyiEE}.

\begin{figure*}[t]
  \includegraphics[width=0.95\textwidth]{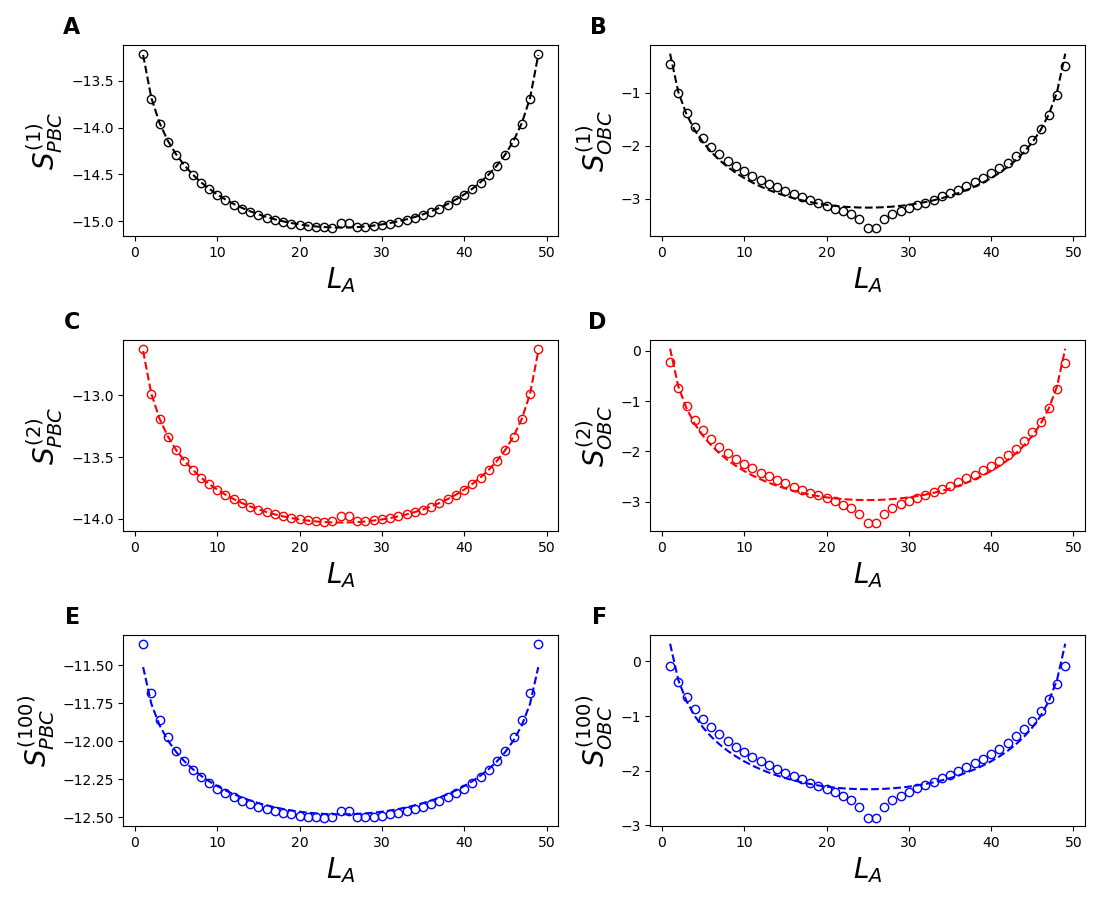}
  \caption{\label{fig:renyiAppen} R\'enyi entropies for PBC (A,C,E) and fined tuned impurity OBC (B,D,F) impurity model \eqref{localperturbation} with $\alpha = 7/5$ and $\beta = 2/5$. Here, $v_{1} = -w_{1} = 1$, and the impurity is placed at $x = 0$. In the PBC case, the n-th order R\'enyi entropy also follows the formula, while for OBC fine-tuned impurity, the higher order $n \geq 2$ R\'enyi entropy doesn't follow the formula,  where (B) $c = -6.303$ (D) $c = -8.707$ (F) $c = -11.450$ by \eqref{RenyiEE}.}
\end{figure*}

\nocite{*}

\end{widetext}

\bibliography{nHRef}
\end{document}